\documentclass[aps,pra,10pt,showpacs,twocolumn,superscriptaddress,longbibliography]{revtex4-1}
\usepackage{graphicx}
\usepackage[usenames]{color}
\usepackage{amssymb,amsmath}
\usepackage{siunitx}
\usepackage{xcolor}
\usepackage{setspace} 
\usepackage{float} 
\usepackage{tabularx}
\usepackage[linkcolor=blue,citecolor=blue,colorlinks=true,urlcolor=blue]{hyperref}
\usepackage[T1]{fontenc}
\usepackage{amsmath,esint}
\usepackage{comment}
\usepackage{lipsum}
\usepackage[toc]{appendix}
\newcommand{\eq}[1]{Eq.~(\ref{#1})}
\newcommand{\fig}[1]{Fig.~\ref{#1}}
\newcommand{\be}[1]{\begin{equation}\label{#1}}
\newcommand{\ee}{\end{equation}}
\begin{document}

\title{Multielectron ionization in O$_2^+$ driven by intense infrared laser pulses}

\author{G. P. Katsoulis}
\affiliation{Department of Physics and Astronomy, University College London, Gower Street, London WC1E 6BT, United Kingdom}
\author{A. Emmanouilidou}
\affiliation{Department of Physics and Astronomy, University College London, Gower Street, London WC1E 6BT, United Kingdom}
\begin{abstract}
We extend a recently developed three-dimensional semiclassical model [\href{https://journals.aps.org/pra/abstract/10.1103/PhysRevA.109.033106}{Phys. Rev. \textbf{A} 109, 033106 (2024)}] to study multielectron ionization and the formation of highly excited Rydberg states in O$_{2}^+$ driven by intense infrared laser pulses. Our model fully accounts for the Coulomb interaction between all particles, except for the Coulomb repulsion between bound electrons which is replaced by effective potentials. This replacement overcomes the hurdle of artificial autoinization. In addition, the multielectron motion is treated on an equal footing with nuclear motion, that is, electrons and nuclei are both allowed to move at the same time. We focus on triple and double ionization as well as frustrated triple and double ionization. For these processes, we identify and explain the main features of the sum of the kinetic energies of the final ion fragments resulting from the break-up of O$_{2}^+$. We  also describe a physical mechanism that underlies frustrated triple ionization. 
\end{abstract}

\date{\today}

\maketitle

\section{Introduction}

Molecules driven by intense infrared laser pulses exhibit a variety of nonlinear processes, such as multielectron ionization which has a central role in attosecond science \cite{RevModPhys.81.163}. Another important process is the formation of highly excited Rydberg states, which have a plethora of applications, such as the acceleration of neutral particles \cite{Eichmann} and the formation of molecules via long-range interactions \cite{Bendkowsky}. It was shown, in the context of  $\mathrm{H_2}$ driven by intense and infrared  laser pulses \cite{Manschwetus,Emmanouilidou_2012}, that formation of Rydberg states proceeds via frustrated ionization. In frustrated ionization, an electron  tunnel-ionizes  due to the driving laser field. However, it does not escape  and is subsequently driven back and recaptured by the parent ion, populating a  Rydberg, i.e. highly excited, state \cite{Nubbemeyer}.

Most studies on strongly driven molecules, i.e. driven by intense infrared laser pulses, address double and frustrated double ionization of  two-electron molecules \cite{Emmanouilidou_2012,PhysRevA.94.043408,Agapi_2016,PhysRevA.96.033404,PhysRevA.101.033403,PhysRevA.108.013120,Zhang2023,Manschwetus,Zhang2,McKenna1, MScKenna2, Sayler,Tiwari2022}.  Diatomic molecules, in particular, offer a unique platform for studying how electron correlation and nuclear motion govern ionization dynamics on the attosecond timescale. Theoretical investigations of three-electron ionization  in strongly driven three-active electron molecules  remain scarce \cite{PhysRevA.103.043109,PhysRevA.109.033106,jfqj-c19k}. This is due to the complexity of molecular strong-field ionization. Indeed, in molecules, the nuclear motion plays an important role and must therefore be treated on an equal footing with electron dynamics. However,  computationally accounting for both multielectron and nuclear motion is a formidable task.

This computational challenge is further evidenced by the limited number of theoretical works addressing three-electron ionization in strongly driven atoms which is still a highly demanding task albeit simpler compared to the study of molecules. Most of the existing studies on atoms, typically formulated within the dipole approximation, rely on reduced-dimensionality classical \cite{PhysRevA.64.053401,PhysRevLett.97.083001} or quantum-mechanical \cite{PhysRevA.98.031401,Efimov:21} models to tackle the computational complexity. However, this reduction in dimensionality leads to an inaccurate description of electron-electron repulsion. As a result, fully three-dimensional (3D) treatments are necessary. At present, 3D models available for studying triple ionization in strongly driven atoms are only classical or semiclassical \cite{PhysRevLett.97.083001,Zhou:10,Tang:13,PhysRevA.104.023113,PhysRevA.105.053119}.

Fully \textit{ab initio} calculations of  triple ionization  in strongly driven molecules are computationally  out of reach for the foreseeable future. These calculations are currently  limited to simpler processes, such as  studying double ionization in $\mathrm{H_2}$ \cite{PhysRevLett.96.143001,PhysRevA.90.053424}. An alternative approach to studying multielectron ionization in intense laser fields is provided by Monte Carlo semiclassical models. A key advantage of these methods is their significantly reduced computational cost compared to quantum mechanical techniques. Moreover, semiclassical models offer valuable insights into the physics underlying  strong-field processes. However, semiclassical models face a hurdle,  artificial autoionization. In contrast to quantum mechanics, classically there is no lower energy limit. As a result, a bound electron can come very close to the core  and acquire large negative energy due to the Coulomb singularity. This energy can then be transferred through Coulomb repulsion to another bound electron, potentially leading to its artificial ionization. To address this issue, most classical and semiclassical models soften the Coulomb potential \cite{PhysRevLett.97.083001,Zhou:10,Tang:13} or introduce Heisenberg-type potentials \cite{PhysRevA.21.834}. The latter  potentials effectively soften the electron-core interaction by mimicing the Heisenberg uncertainty principle and preventing close encounters between electrons and the core \cite{PhysRevA.104.023113,PhysRevA.105.053119}. However, softening the Coulomb potential does not allow for an accurate description of electron scattering from the core \cite{Pandit2018,Pandit2017}, leading to ionization spectra that can significantly deviate from experimental results, as demonstrated, for example, in driven Ne and Ar \cite{Agapi3electron,PhysRevA.107.L041101}.

Recently, we developed a three-dimensional semiclassical model to study multielectron ionization and fragmentation of molecules driven by strong laser fields while treating the autoinization problem \cite{PhysRevA.109.033106}.  Specifically, the model accounts exactly for the Coulomb interactions between all particles, except for the Coulomb interaction between bound electrons. Specifically, we replace the Coulomb repulsion by employing effective Coulomb potentials to describe the bound-bound electron interaction, hence the acronym ECBB for our model. During time propagation, we identify on the fly whether electrons are bound or quasifree and activate the  full  or effective Coulomb interaction accordingly \cite{PhysRevA.109.033106}. The ECBB model for molecules  generalized the ECBB model we have previously developed for strongly driven atoms \cite{Agapi3electron,PhysRevA.108.043111,PhysRevA.107.L041101,43wt-x129,Praill_2026}. Using the ECBB model, we studied triple and double ionization, as well as the  formation of Rydberg states in the strongly driven triatomic  $\mathrm{HeH_2^+}$  \cite{PhysRevA.109.033106}.

Here, we extend our recently developed 3D ECBB model for molecules to study strong-field multielectron ionization of O$_2^+$. Specifically, for the initial conditions of the tunnel-ionizing electron, we account for the Coulomb interaction between this electron and the bound ones by integrating over the electronic density of the bound electrons.  In previous studies, see Refs. \cite{toolkit2014,PhysRevA.103.043109,PhysRevA.109.033106}, the wave function in the electronic density of the bound electrons is expressed in terms of $s$-symmetry Gaussian functions. In this work, to account for the Coulomb interaction, we  perform more involved integrations using an electronic density expressed in terms of higher than $s$-symmetry wave functions, relevant to O$_{2}^{+}$.
 In addition, we modify tunneling during time propagation to incorporate the effective potentials used to describe the interaction between bound electrons. Using the  ECBB model, we investigate triple and double ionization as well as frustrated triple and double ionization. For these processes,  we obtain the kinetic energy release (KER), i.e. the sum of the kinetic energies of the ion fragments. We find that these KER have larger values  compared to experiment \cite{PhysRevA.79.063414}. 
 However, when the momentum change due to the effective potential is not accounted for, we find good agreement with experiment.  This allows us to understand the limitations of the ECBB model and for which molecules our results for KER will be closer to experiment. Finally, we also find
 the main mechanisms of TI, FDI and FTI.
\section{Method}\label{Sec::Method}

In what follows, we briefly describe the ECBB model in the context of  strongly driven O$_2^+$, which is modelled as a three-active-electron molecule. Details of the theoretical framework of the ECBB model are described in Ref. \cite{PhysRevA.109.033106}. We also discuss in detail how we extend the ECBB model  described in Ref. \cite{PhysRevA.109.033106} to account for the O$_2^+$ molecule.

We model the driving laser field using the vector potential, $\mathbf{A}(y,t)$, of the form
\begin{equation}\label{eq:vector_potential}
\mathbf{A}(\mathrm{y,t}) = -\frac{E_0}{\omega}\exp \left[ - 2\ln (2)\left( \frac{\mathrm{c t - y}}{\mathrm{c} \tau} \right)^2 \right]   \sin ( \omega \mathrm{t}  - \mathrm{k y}) \hat{\mathbf{z}},
\end{equation}
where $E_0$ is the field strength, $\omega$ is the radial frequency, $\mathrm{k=\omega/c}$ is the wave number of the laser field and  $\tau$ is the full width at half maximum of the pulse duration in intensity. The direction of both the vector potential and the electric field, $\mathbf{E}(y,t) = - \dfrac{\partial\mathbf{A}(y,t)}{\partial t}$, is along the $z$ axis. We take the propagation direction of the laser field to be along the $y$ axis and hence the magnetic field points  along the $x$ axis. We note that the ECBB model fully accounts for the magnetic field of the laser field, however, we do not address magnetic-field effects in this work.

We start the propagation at  time $t_0$, which is found using importance sampling \cite{ROTA1986123} in the time interval $[-2\tau , 2\tau ]$ where the electric field is nonzero. The sampling distribution  is the ionization rate, which is obtained as described in detail in Appendix \ref{Sec::ionrate}. Then we specify the initial conditions of all particles.  One of the electrons escapes from the molecule, either via tunnelling or via over-the-barrier ionization, see Sec. \ref{Sec::Exit_poin_of_e1}. The remaining two bound electrons are initialized using a microcanonical distribution, see Sec. \ref{Sec::microcanonical_distribution}. Both nuclei are placed along the $z$ axis, at -$R(t_0)/2$ and $R(t_0)/2$ respectively,  with the origin of the coordinate system set to be the center of mass of the molecule. We compute the internuclear distance at time $t_0,$ $R(t_0)=2.114$ a.u., and the first and second ionization energies, $I_{p1}, I_{p2}$ using the quantum chemistry package MOLPRO \cite{MOLPRO_brief}. Both nuclei are initiated at rest.

 \begin{figure}[t]
\centering
\includegraphics[width=\columnwidth]{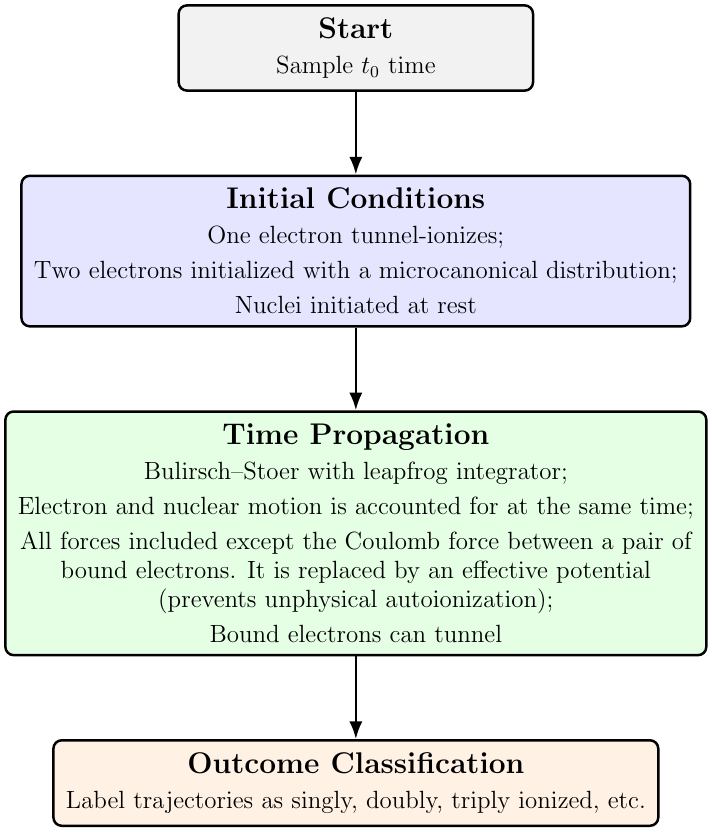}
\caption{Schematic illustration of the main steps involved in the ECBB model.}\label{Fig:ECBB_model}
\end{figure}
 
 The subsequent time evolution is performed using a leapfrog technique \cite{Pihajoki2015,toolkit2014,PhysRevA.103.033115} jointly with the Bulirsch-Stoer method \cite{press2007numerical,bulirsch1966numerical}. Details can be found in Ref. \cite{PhysRevA.109.033106}. During time propagation all particles are treated on an equal footing, i.e., both the electrons and the nuclei are allowed to move at the same time  as dictated by  the nondipole Hamiltonian that fully accounts for the magnetic-field component of the laser field. Also, all Coulomb interactions are included explicitly, except for the interaction between a pair of bound electrons, which is replaced by an effective potential to prevent unphysical autoionization, see Sec. \ref{Sec::Hamiltonian}. Finally, in addition to the tunneling rate, we incorporate another quantum aspect in our model. Namely, during time propagation, we allow for each bound electron to tunnel, see Sec. \ref{Sec::Tunnelling}.  We stop the time propagation when the energy of each particle converges. Then, we label the trajectory as triply or doubly ionized if three or two electrons, respectively, have positive energies. The triple ionization (TI) and double ionization (DI) probabilities are computed out of all events. In \fig{Fig:ECBB_model} we present a schematic illustration of the main steps involved in the ECBB model.

\subsection{Hamiltonian of the system}\label{Sec::Hamiltonian}
We treat O$_{2}^{+}$ as a three-active-electron molecule. The relevant Hamiltonian of the 5-body system, comprised of 2 nuclei (indexes 1 and 2) and 3 electrons (indexes 3, 4 and 5) in the nondipole approximation is given by
{\allowdisplaybreaks
\begin{align}\label{eq:Ham}
\begin{split}
&H = \sum_{i=1}^{5}\frac{\left[\mathbf{\tilde{p}}_{i}- Q_i\mathbf{A}(\mathbf{r}_{i},t) \right]^2}{2m_i}+\sum_{n=1}^{2}\sum_{j= n +1}^{5}\frac{Q_nQ_j}{|\mathbf{r}_n-\mathbf{r}_j|}  \\
&+\sum_{i=3}^{4}\sum_{j=i+1}^{5} \left[ 1-c_{i,j}(t)\right]\frac{Q_i Q_j}{|\mathbf{r}_i-\mathbf{r}_j|} +\sum_{i=3}^{4}\sum_{j=i+1}^{5}c_{i,j}(t)V_{i,j}\\
\end{split}
\end{align}}

\noindent where ${Q_i}$ is the charge, ${m_i}$ is the mass, $\mathbf{r}_{{i}}$ is the position vector and $\mathbf{\tilde{p}}_{{i}}$ is the canonical momentum vector of particle $i$. The mechanical momentum ${\mathbf{p}_{i}}$ is given by
\begin{equation}\label{eq:mechanical_momentum}
{\mathbf{p}_{i}=\mathbf{\tilde{p}}_{{i}}- {Q_i}\mathbf{A}(\mathbf{r}_{{i}},{t}}).
\end{equation}
The effective Coulomb potential that an electron $i$ experiences at a distance $|\mathbf{r}_{n}-\mathbf{r}_{i}|$ from the nucleus $n$, due to the charge distribution of electron $j$ is given by \cite{Agapi3electron,PhysRevA.40.6223}
\begin{align}\label{eq:eff_potential}
\begin{split}
&V_{\text{eff}}(\zeta_{j,n}(t),|\mathbf{r}_{n}-\mathbf{r}_{i}|) \\
& = \frac{1 - (1+\zeta_{j,n}|\mathbf{r}_{n}-\mathbf{r}_{i}|)e^{-2\zeta_{j,n}|\mathbf{r}_{n}-\mathbf{r}_{i}|}}{|\mathbf{r}_{n}-\mathbf{r}_{i}|},
\end{split}
\end{align}
with $\zeta_{j,n}$ the effective charge of electron $j$ \cite{Agapi3electron,PhysRevA.40.6223}.
We note that, when electron $i$ undergoes a close approach with one of the nuclei, i.e., $ \mathbf{r}_{i}\rightarrow\mathbf{r}_{n}$, the effective potential is equal to $\zeta_{j,n}$, hence, ensuring a finite energy transfer between bound electrons $i$ and $j$. As a result, no artificial autoionization takes place.

The effective charge $\zeta_{j,n}(t)$ of particle $j,$ at time $t$ is defined as
\begin{equation}\label{eqn::zeta_and_energy}
\zeta_{j,n}(t) = \left\{
    \begin{array}{ll}
        Q_n & \mathcal{E}_{j}(t)  \leq \mathcal{E}_{{1 s}}\\
        \left(Q_n / \mathcal{E}_{{1 s}}\right) \mathcal{E}_{j}(t) & \mathcal{E}_{{1 s}} < \mathcal{E}_{j}(t)  < 0\\
        0 & \mathcal{E}_{j}(t)  \geq 0,
    \end{array}
\right.
\end{equation}
where $Q_n$ is the charge of the nucleus $n$, $\mathcal{E}_{{1 s}}$ is the sum of the ground-state energies of a hydrogenic atom with nuclear charge $Q_n$, i.e., $\mathcal{E}_{{1 s}} = \sum_{n=1}^{2} -Q^2_n/2,$ and  $\mathcal{E}_{j}(t)$ is the energy of electron $j$ given by 
{\allowdisplaybreaks
\begin{align}\label{eq:energy_of_electron_j}
\begin{split}
\mathcal{E}_{j}(t) &= \frac{\left[\mathbf{\tilde{p}}_{j}- Q_j\mathbf{A}(\mathbf{r}_{j},t) \right]^2}{2m_j} + \sum_{n=1}^{2}\frac{Q_n Q_j}{|\mathbf{r}_{n}-\mathbf{r}_{j}|}  \\ 
&+ \sum_{n=1}^{2}\sum_{\substack{\;{i=3} \\ {i} \neq {j}}}^{5} c_{i,j}(t)C_{i,n}(\mathcal{E}_{i},|\mathbf{r}_{1}-\mathbf{r}_{i}|,|\mathbf{r}_{2}-\mathbf{r}_{i}|)\\
&\times V_{\text{eff}}(\zeta_{i,n}(t),|\mathbf{r}_{n}-\mathbf{r}_{j}|) - Q_j\mathbf{r}_{j} \cdot \mathbf{E}\left(\mathbf{r}_{j}, t\right).
\end{split}
\end{align}}
The definition of $\mathcal{E}_{1 s} $ as the sum of hydrogenic energies, $\sum_{n=1}^{2} -Q^2_n/2,$ originates from  $\mathcal{E}_{j}(t)$ being expressed in terms that involve all nuclei. Indeed, although the effective charge $\zeta_{j,n}(t)$ is associated with the nucleus $n$, the energy of electron $j$ contains terms, e.g., the kinetic energy, that cannot be partitioned between the two nuclei. 

The potential $V_{i,j}$ in the Hamiltonian \eqref{eq:Ham} is given as a sum of effective potentials as follows
{\allowdisplaybreaks
\begin{align}
\begin{split}
&V_{i,j} = \sum_{n=1}^{2} \Big[ C_{j,n}(\mathcal{E}_{j},|\mathbf{r}_{1}-\mathbf{r}_{j}|,|\mathbf{r}_{2}-\mathbf{r}_{j}|) V_{\text{eff}}(\zeta_{j,n}(t),|\mathbf{r}_{n}-\mathbf{r}_{i}|) \nonumber \\
&  + C_{i,n}(\mathcal{E}_{i},|\mathbf{r}_{1}-\mathbf{r}_{i}|,|\mathbf{r}_{2}-\mathbf{r}_{i}|) V_{\text{eff}}(\zeta_{i,n}(t),|\mathbf{r}_{n}-\mathbf{r}_{j}|) \Big],
\end{split}
\end{align}}

\noindent where  $C_{i,n}(\mathcal{E}_{i},|\mathbf{r}_{1}-\mathbf{r}_{i}|,|\mathbf{r}_{2}-\mathbf{r}_{i}|) $ is the probability distribution $\rho_{i,n}$ for an electron $i$ to be found around nucleus  $n$, as if there are no other nuclei, and then dividing by the sum of all such distributions from all different nuclei
\begin{equation}\label{Cim_constant}
C_{i,n}(\mathcal{E}_{i},|\mathbf{r}_{1}-\mathbf{r}_{i}|,|\mathbf{r}_{2}-\mathbf{r}_{i}|)  = \frac{\rho_{i,n}}{{\sum_{{n'=1}}^{2}{\rho_{i,n'}}}} ,
\end{equation}
with
\begin{align}\label{unnorm_dense}
\begin{split}
{\rho_{i,n}(t)} &= | \psi (\zeta_i,|\mathbf{r}_{n}-\mathbf{r}_{i}|) |^2 = {\frac{\zeta_{i,n}^3}{\pi}e^{-2\zeta_{i,n}(t)|\mathbf{r}_{n}-\mathbf{r}_{i}|}},
\end{split}
\end{align}
where  we approximate the wave function of each bound electron $i$ with a $1s$ hydrogenic wavefunction. 

At every time step during propagation, the functions ${c_{i,j}(t)}$ determine whether the interaction between electrons $i,j$ is described by the full Coulomb potential or by the effective potentials $V_{\text{eff}}(\zeta_{i},|\mathbf{r}_{n}-\mathbf{r}_{j}|)$ and $V_{\text{eff}}(\zeta_{j},|\mathbf{r}_{n}-\mathbf{r}_{i}|)$. The exact mathematical definition of ${c_{i,j}(t)}$ is given in Ref. \cite{PhysRevA.109.033106}, while ${c_{i,j}(t)}=0$  corresponds to the full Coulomb potential and ${c_{i,j}(t)}=1$  corresponds to the effective potential. The effective potentials are employed only when both electrons in the pair $(i,j)$ are classified as bound. If at least one of the electrons is quasifree, we use the full Coulomb potential. We determine on the fly whether an electron is bound or quasifree during time propagation. To make this determination, we use the following criteria. A quasifree electron may become bound after a recollision. Specifically, once it reaches its point of closest approach to a nucleus, it is labelled as bound if its subsequent motion along the $z$ axis is governed more strongly by the Coulomb attraction potential than by the laser field. Conversely, a bound electron may transition to quasifree either through energy transfer during a recollision or via the laser field. In the former case, the transition occurs when the electron's potential energy with respect to the nucleus decreases constantly. In the latter case, if the electron's compensated energy \cite{Leopold_1979} becomes positive and remains positive, it is deemed quasifree. A detailed discussion and illustrative examples of these criteria are provided in Ref. \cite{Agapi3electron}. We note, that at time $t_0$, i.e., at the start of the time propagation, one of the electrons (index 5) tunnel-ionizes while the other two (indexes 3 and 4) are bound. Therefore, we have $c_{3,4}(t_0)=1$ and $c_{3,5}(t_0)=c_{4,5}(t_0)=0.$

\subsection{Initial conditions for the electrons}
\subsubsection{Tunnel-ionizing electron}\label{Sec::Exit_poin_of_e1}
At the start of the propagation $t_{0}$, one electron (denoted by the index 5) tunnel-ionizes  through the  Coulomb potential barrier that is lowered by the electric field.
This is the case if the field strength  is within the below-the-barrier ionization regime.  To find the exit point, we use the equation
\begin{align}\label{eq:App3}
\begin{split}
&V({r}_{5,\parallel},t)=-\frac{Q_1}{\left\vert\mathbf{r}_{1}-\mathbf{r}_{5}\right \vert}-\frac{Q_2}{\left \vert\mathbf{r}_{2}-\mathbf{r}_5\right \vert} \\
&+\sum_{i=3}^{4}\int\frac{\vert\Psi(\mathbf{r}_i)\vert^2}{\vert\mathbf{r}_5-\mathbf{r}_i\vert}\textrm{d}\mathbf{r}_i +\mathbf{r}_5\cdot\mathbf{E}(t) = -I_{p1},
\end{split}
\end{align} 
with $Q_1,Q_2$ the charges of the two nuclei, equal to 2 a.u. in this work, and $\mathbf{r}_{1},\mathbf{r}_{2}$ are the position vectors of the nuclei. The position vector of the tunnel-ionizing electron is $\mathbf{r}_{5}$. We specify the initial conditions within the dipole approximation, i.e., we ignore the spatial dependence of the electric field, $\mathbf{E}(t) \equiv \mathbf{E}(\mathbf{0},t).$ As already stated in the Introduction, in the extended ECBB model,  the wave function, $\Psi(\mathbf{r}_i)$, is expanded in a basis of Gaussian functions with higher than $s$-symmetry. In contrast, in our previous work, see Refs. \cite{toolkit2014,PhysRevA.103.043109,PhysRevA.109.033106}, only up to $s$-symmetry was considered. We solve \eq{eq:App3} for $r_{5,\parallel}$, i.e., the component of $\mathbf{r}_{5}$ along the direction of the electric field, while setting the component of $\mathbf{r}_{5}$ perpendicular to the field equal to zero.  The integral, $\int\frac{\vert\Psi(\mathbf{r}_i)\vert^2}{\vert\mathbf{r}_5-\mathbf{r}_i\vert}\textrm{d}\mathbf{r}_i$, accounts for the potential of the tunnel-ionizing electron due to the charge distribution of bound electron $i$.  In the current work, we model O$_2^+$ as a 3 active electron molecule. The electronic configuration of the ground state of O$_2$ is
\begin{equation}
\left(1 \sigma_g^2, 1 \sigma_u^2, 2 \sigma_g^2, 2 \sigma_u^2, 3 \sigma_g^2, 1 \pi_{u x}^2, 1 \pi_{u y}^2, 1 \pi_{g x}^1, 1 \pi_{g y}^1 \right),
\end{equation}
with the two outermost orbitals being degenerate. Therefore, for our semiclassical calculations it will not make a difference whether one electron is missing from the $1 \pi_{g x}^1$ or the $1 \pi_{g y}^1$ orbital of O$_2$. We chose the later, leading to the electronic configuration of the ground state of O$_2^+$ being
\begin{equation}
\left(1 \sigma_g^2, 1 \sigma_u^2, 2 \sigma_g^2, 2 \sigma_u^2, 3 \sigma_g^2, 1 \pi_{u x}^2, 1 \pi_{u y}^2, 1 \pi_{g x}^1 \right).
\end{equation}
Using MOLPRO, we found that the energy of the outermost orbital $1 \pi_{g x}^1$ is smaller, therefore, one electron tunnels out from the $1 \pi_{g x}^1$ orbital while the two bound electrons occupy the $ 1\pi_{u y}^2$ of O$_2^{2+}$. We note that the two remaining electrons could have occupied the $ 1\pi_{u x}^2$ orbital, however, the calculations would have been equivalent due to the degeneracy of these two orbitals.
The sum in Eq. \eqref{eq:App3} takes the form
\begin{equation}\label{Eq:KeeO2py}
\sum_{i=3}^{4}\int\frac{\vert\Psi(\mathbf{r}_i)\vert^2}{\vert\mathbf{r}_5-\mathbf{r}_i\vert}\textrm{d}\mathbf{r}_i = 2 \int \frac{\left|\Psi^{1\pi_{u y}}\left(\mathbf{r}\right)\right|^2}{\left|\mathbf{r}_5-\mathbf{r}\right|} \mathrm{d} \mathbf{r},
\end{equation}
where for simplicity we dropped the subscript $i$ from the position vector of electron $i$, i.e., $\mathbf{r} \equiv \mathbf{r}_i.$ To calculate this integral, we express the wavefunction $\Psi^{1\pi_{u y}}(\mathbf{r})$ in terms of $N_{\text {contr }}$ contracted Gaussians, i.e.,
\begin{equation}\label{Eq:Wavefunction}
\Psi^{1\pi_{u y}}(\mathbf{r})=\sum_{j=1}^{2} \sum_{n=1}^{N_{\text {contr }}} c_{j, n} \phi_{j, n}\left(\mathbf{r}-\mathbf{r}_j\right)
\end{equation}
where the $n$-contracted Gaussian for the $j$ nucleus with position vector $\mathbf{r}_j$ reads
\begin{equation}\label{Eq:contracted}
\phi_{j, n}\left(\mathbf{r}-\mathbf{r}_j\right)=\sum_{i=1}^{m_n} \sum_{k=a}^{c} d_{j, n, i, k} G\left(\mathbf{r}-\mathbf{r}_j, \alpha_{j, n, i, k}\right),
\end{equation}
with $m_n$  the number of primitives for the $n$-contracted Gaussian and $d_{j, n, i, k}$ the contraction coefficients. The index $k$ sums over all the combinations of the $a,b,c$ in \eq{Eq:primitives} that correspond to the same symmetry of the primitive. The Gaussian type primitives are of the form,
\begin{equation}\label{Eq:primitives}
G(x, y, z ; \alpha_{j, n, i, k})=\mathcal{N}_{a b c ; \alpha_{j, n, i, k} } x^a y^b z^c e^{ -\alpha_{j, n, i, k}\left(x^2+y^2+z^2\right)},
\end{equation}
with
\begin{equation}
\mathcal{N}_{a b c ; \alpha_{j, n, i, k} }=\left(\frac{2 \alpha_{j, n, i, k}}{\pi}\right)^{3 / 4}\left[\frac{(8 \alpha_{j, n, i, k})^{a+b+c} a!b!c!}{(2 a)!(2 b)!(2 c)!}\right]^{1 / 2}
\end{equation}
a normalization constant. Hence, the integral in \eq{Eq:KeeO2py} takes the form,
\begin{align}\label{eq::integral}
\begin{split}
&\int \frac{\left|\Psi^{1\pi_{u y}}\left(\mathbf{r}\right)\right|^2}{\left|\mathbf{r}_5-\mathbf{r}\right|} \mathrm{d} \mathbf{r} = \sum_{j, j^{\prime}} \sum_{n, n^{\prime}} \sum_{i, i^{\prime}} \sum_{k, k^{\prime}} c_{j, n} c_{j^{\prime}, n^{\prime}}  \\
&\times d_{j, n, i, k} d_{j^{\prime}, n^{\prime}, i^{\prime}, k^{\prime}} I\left(\mathbf{r}_5, \mathbf{r}_j, \mathbf{r}_{j^{\prime}}, \alpha_{j, n, i, k}, \alpha_{j^{\prime}, n^{\prime}, i^{\prime},  k^{\prime}}\right),
\end{split}
\end{align}
where we set the integral to be
\allowdisplaybreaks{
\begin{align}\label{eq::integral_v2}
\begin{split}
&I\left(\mathbf{r}_5, \mathbf{r}_j, \mathbf{r}_{j^{\prime}}, \alpha_{j, n, i, k}, \alpha_{j^{\prime}, n^{\prime}, i^{\prime}, k^{\prime}}\right) = \\
&\int  \frac{G\left(\mathbf{r}-\mathbf{r}_j, \alpha_{j, n, i, k}\right) G\left(\mathbf{r} -\mathbf{r}_{j^{\prime}}, \alpha_{j^{\prime}, n^{\prime}, i^{\prime}, k^{\prime}}\right)}{\left|\mathbf{r}_5-\mathbf{r} \right|}\mathrm{d} \mathbf{r}.
\end{split}
\end{align}}
We obtain the coefficients $c_{j,n}$, $d_{j,n,i,k}$ and $\alpha_{j,n,i,k}$  from a Hartree-Fock calculation with MOLPRO using the aug-cc-pVQZ basis set. The calculation of the integral in \eq{eq::integral_v2} is performed analytically if the symmetry of the Gaussian functions allows so, otherwise we solve the integral numerically. Specifically, when summing over s-symmetry Gaussian functions we find the integral to be \cite{toolkit2014}
\begin{align}
\begin{split}
I\left(\mathbf{r}_5, \mathbf{r}_j, \mathbf{r}_{j^{\prime}}, \alpha, \beta\right)= & \frac{(4 \alpha \beta)^{3 / 4}}{(\alpha+\beta)^{3 / 2}} \frac{\operatorname{erf}\left(\sqrt{\alpha+\beta}\left|\mathbf{r}_5-\mathbf{ \tilde{r} }\right|\right)}{\left|\mathbf{r}_5-\mathbf{ \tilde{r} }\right|} \\
& \times \exp \left[-\frac{\alpha \beta\left(\mathbf{r}_j-\mathbf{r}_{j^{\prime}}\right)^2}{\alpha+\beta}\right]
\end{split}
\end{align}
with $\mathbf{ \tilde{r} } = \left(\alpha \mathbf{r}_j+\beta \mathbf{r}_{j^{\prime}}\right) /(\alpha+\beta)$ and $\operatorname{erf}(x)$ the error function \cite{abramowitz1948handbook}. Moreover, for simplicity we dropped the indexes $j,n,i,k$ and $j^{\prime}, n^{\prime}, i^{\prime}, k^{\prime}$, that is, $\alpha \equiv \alpha_{j,n,i,k} $ and $\beta \equiv \alpha_{j^{\prime}, n^{\prime}, i^{\prime}, k^{\prime}}$.
For the rest of the symmetries of the Gaussian functions we find the integral to be 
\allowdisplaybreaks{
\begin{align}\label{Eq:Int_general_final}
&I\left(\mathbf{r}_5, \mathbf{r}_j, \mathbf{r}_{j^{\prime}}, \alpha, \beta\right) =\dfrac{2\pi }{\alpha + \beta } N_{ k} N_{ k^{\prime} }   \nonumber \\
&\times \exp \left[-\frac{\alpha \beta\left(\mathbf{r}_j-\mathbf{r}_{j^{\prime}}\right)^2}{\alpha+\beta}\right] \sum_{k_a=0}^{a} \sum_{k^{\prime}_a=0}^{a^{\prime}} \sum_{k_b=0}^{b} \sum_{k^{\prime}_b=0}^{b^{\prime}} \sum_{k_c=0}^{c} \sum_{k^{\prime}_c=0}^{c^{\prime}} \nonumber  \\
&\times \binom{a}{k_a}\binom{a^{\prime}}{k^{\prime}_a}\binom{b}{k_b}\binom{b^{\prime}}{k^{\prime}_b} \binom{c}{k_c}\binom{c^{\prime}}{k^{\prime}_c} \left[ \left(k_a+k^{\prime}_a-1\right)!! \right] \nonumber  \\
&\times \left[ \left(k_b+k^{\prime}_b-1\right)!! \right] \left[ \left(k_c+k^{\prime}_c-1\right)!! \right]  \int_{0}^{1} e^{-(\alpha +\beta )t^2| \mathbf{ \tilde{r} } -\mathbf{r}_5 |^2} \nonumber  \\
&\times \left[ \frac{1-t^2}{2(\alpha +\beta)} \right]^{\left(k_a+k^{\prime}_a+k_b+k^{\prime}_b+k_c+k^{\prime}_c\right) / 2}  \nonumber   \\
&\times \left[\tilde{r}_x -t^2(\tilde{r}_x-x_5) - x_j \right]^{a-k_a }  \nonumber  \\
& \times \left[\tilde{r}_x -t^2(\tilde{r}_x-x_5)-x_{j^{\prime}}\right]^{a^{\prime}-k^{\prime}_a} \nonumber  \\
&\times \left[\tilde{r}_y -t^2(\tilde{r}_y-y_5) - y_j \right]^{b-k_b} \nonumber  \\
&\times \left[\tilde{r}_y -t^2(\tilde{r}_y-y_5)-y_{j^{\prime}}\right]^{b^{\prime}-k^{\prime}_b} \nonumber  \\ 
&\times \left[\tilde{r}_z -t^2(\tilde{r}_z-z_5) - z_j \right]^{c-k_c} \nonumber  \\
&\times \left[\tilde{r}_z -t^2(\tilde{r}_z-z_5)-z_{j^{\prime}}\right]^{c^{\prime}-k^{\prime}_c}  \mathrm{d}t,
\end{align}}
where we expressed the relevant position vectors into their respective coordinates. That is, $\mathbf{r}_5=(x_5,y_5,z_5)$, $\mathbf{r}_j=(x_j,y_j,z_j)$, $\mathbf{r}_j^{\prime}=(x_{j^{\prime}},y_{j^{\prime}},z_{j^{\prime}})$ and $\mathbf{ \tilde{r} }= (\tilde{r}_x,\tilde{r}_y,\tilde{r}_z)$. Moreover, the subscripts $k,k^{\prime}$ in the normalization constants $N_{k}$  and  $N_{k^{\prime}} $ denote the different exponents $a,b,c$ and  $a^{\prime},b^{\prime},c^{\prime}$ [see \eq{Eq:primitives}] corresponding to the Gaussian primitives with exponents $\alpha \equiv \alpha_{j,n,i,k}$ and $\beta \equiv \alpha_{j^{\prime}, n^{\prime}, i^{\prime}, k^{\prime}}$ respectively. We note that the numerical evaluation of the integral in \eq{Eq:Int_general_final}  is considerably less demanding than computing the integral in \eq{eq::integral_v2}. This reduction in computational cost arises because the transformations we performed reduced the integral to a one-dimensional integral over a finite domain [0,1], rather than a three-dimensional integral over an infinite region. Hence, although the resulting expression is not analytical, the integral can be computed efficiently. 

Conserning  the momentum of the tunnel-ionizing electron, we assume that it exits the field-lowered Coulomb barrier with a zero momentum along the direction of the field. The transverse electron momentum is given by a Gaussian distribution. The latter arises from standard tunneling theory \cite{Delone:91,Delone_1998,PhysRevLett.112.213001} and represents the Gaussian-shaped filter with an intensity-dependent width. 

If the field strength of the electric field at time $t_0$ is larger than the threshold intensity for over-the-barrier ionization, then the electron is placed at the top of the barrier. The electron exits opposite to the field direction at a  distance $r_{\text {max }}$, with $r_{\text {max }}$ being the coordinate along the electric field axis where the field-lowered Coulomb potential $V\left(r_{5, \|}, t_0\right)$  has its maximum. The magnitude of the initial momentum of the escaping electron is then taken as
\begin{equation}
{p}_5=\sqrt{2\left[(-I_{p1})-V\left(r_{\max }, t_0\right)\right]}.
\end{equation}
The direction of $\mathbf{p}_5$ is sampled randomly in space, with the only restriction being that $\mathbf{p}_5 \cdot \mathbf{E}\left(t_0\right) \leqslant 0$. This ensures that the electron moves opposite to the instantaneous field.

\subsubsection{Microcanonical distribution}\label{Sec::microcanonical_distribution}
\begin{figure}[b]
\centering
\includegraphics[width=\columnwidth]{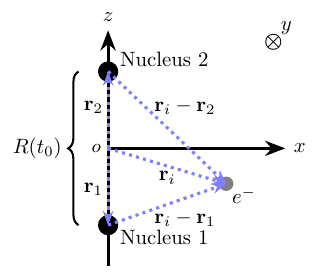}
\caption{The configuration of the diatomic molecule we use to set-up the microcanonical distribution. The origin of the coordinate system is halfway between nuclei 1 and 2.}
\label{Fig:Molecule_microcanonical}
\end{figure}

For the remaining two bound electrons, we obtain their initial position and momentum by  employing a simplified version of the microcanonical distribution described in detail in Ref. \cite{PhysRevA.109.033106} for the triatomic molecule $\mathrm{HeH_2^{+}}$ and in Ref. \cite{Chen_2016} for $\mathrm{H_3^{+}}$. Specifically, the energy of each bound electron $i$ (with $i=3,4$) at time $t_0$ is equal to 
\begin{equation}\label{eq:potential_energy_of_electron_ioi}
\mathcal{E}_i(t_0) = \frac{\mathbf{{p}}_{i}^2}{2m_i} + W_i,
\end{equation}
where 
\begin{align}\label{eq:potential_energy_of_electron_i}
\begin{split}
W_i &= \sum_{n=1}^{2}\frac{Q_n Q_i}{|\mathbf{r}_{n}-\mathbf{r}_{i}|}  + \sum_{n=1}^{2}\sum_{\substack{\;{j=3} \\ {i} \neq {j}}}^{4} c_{i,j}(t_0) \\
&\times C_{j,n}(\mathcal{E}_j,|\mathbf{r}_{1}-\mathbf{r}_{j}|,|\mathbf{r}_{2}-\mathbf{r}_{j}|) V_{\text{eff}}(\zeta_{j,n}(t_0),|\mathbf{r}_{n}-\mathbf{r}_{i}|).
\end{split}
\end{align}
We note that  the summation in \eq{eq:potential_energy_of_electron_i} is  only over the bound electrons. That is, we consider the microcanonical distribution in the O$_{2}^{2+}$ molecule, since electron 5 has tunnel-ionized in the O$_{2}^{+}$ molecule.  We take the energy of each bound electron to be equal to $-I_{p2},$ with $I_{p2}$ being the second ionization potential of the molecule under consideration. We find the effective charges of the bound electrons, $\zeta_{j,n}(t_0),$ using \eq{eqn::zeta_and_energy} with $\mathcal{E}_j(t_0)=-I_{p2}$. Finally, we note that the potential energy $W_i$ of each electron $i$ depends on the coordinates of both bound electrons, due to the function  $C_{j,n}(\mathcal{E}_j,|\mathbf{r}_{1}-\mathbf{r}_{j}|,|\mathbf{r}_{2}-\mathbf{r}_{j}|)$. This  leads to the microcanonical distributions of both bound electrons being interrelated as follows
\begin{equation}\label{eq:multie_micro}
f(\mathbf{r}_3, \mathbf{p}_3, \mathbf{r}_4, \mathbf{p}_4)=\mathcal{N} \prod_{i=3}^{4} \delta\left[\frac{p_i^2}{2}+W_i- (-I_{p2})\right],
\end{equation}
where $N$ is a normalisation constant.

To set-up the microcanonical distribution, we place the origin of the coordinate system  halfway between nuclei 1 and 2, see \fig{Fig:Molecule_microcanonical}. Moreover, $\lambda_i,\mu_i$ are the confocal elliptical coordinates of electron $i$ defined using two of the nuclei as the foci of the ellipse with $\lambda_i \in [1,\infty)$  and  $\mu_i \in [-1,1]$ and are given by 
\begin{align}
\lambda_i &= \frac{ \vert \mathbf{r}_i - \mathbf{r}_1 \vert + \vert \mathbf{r}_i - \mathbf{r}_2 \vert }{R(t_0)}, \\
\mu_i &= \frac{ \vert \mathbf{r}_i - \mathbf{r}_1 \vert - \vert \mathbf{r}_i - \mathbf{r}_2 \vert }{R(t_0)},
\end{align}
with $\vert \mathbf{r}_i - \mathbf{r}_1 \vert$ and $\vert \mathbf{r}_i - \mathbf{r}_2 \vert$ the magnitude of the relative position between electron $i$ and the two nuclei and $R(t_0)$ the distance between the two nuclei, see \fig{Fig:Molecule_microcanonical}.  The coordinate $\phi$  is the angle between the projection of the position vector $\mathbf{r}_i $ of the bound electron $i$ on the $xy$ plane and the positive $x$ axis. The $z$ axis goes through the two nuclei that define the elliptical coordinates. Hence, the angle $\phi$ defines the rotation angle around the $z$ axis, for more details see \cite{Agapi_2016}. Transforming from Cartesian to elliptical coordinates, we find that the microcanonical distribution has the following form
\begin{align}\label{eq:multie_micro2}
\begin{split}
&\rho\left(\lambda_3, \mu_3, \phi_3,\lambda_4, \mu_4, \phi_4\right)=\mathcal{N}'\left(\frac{R^{3}_{ab}}{8}\right)^{2} \prod_{i=4}^{5} \left(\lambda_i^2-\mu_i^2\right)\\
&\times  \sqrt{2(E_i-W_i(\lambda_3, \mu_3, \phi_3,\lambda_4, \mu_4, \phi_4)} \delta\left(E_i+I_{p2}\right),
\end{split}
\end{align}
with the potential energy equal to 
\begin{align} \label{eq:potential_energy_of_electron_i_ell}
\begin{split}
&W_i = \sum_{n=1}^{2}\frac{Q_iQ_n}{|\mathbf{r}_{n}-\mathbf{r}_{i}|}+ \sum_{n=1}^{2}\sum_{\substack{\;{j=3} \\ {j} \neq {i}}}^{4} c_{i,j}(t_0) \\
&\times C_{j,n}(\mathcal{E}_{j},|\mathbf{r}_{1}-\mathbf{r}_{j}|,|\mathbf{r}_{2}-\mathbf{r}_{j}|) V_{\text{eff}}(\zeta_{j,n}(t_0),|\mathbf{r}_{n}-\mathbf{r}_{i}|) 
\end{split}
\end{align}
where
\begin{align}
|\mathbf{r}_{1}-\mathbf{r}_{i}| &= \frac{R(t_0) (\lambda_i + \mu_i)}{2}  \\
|\mathbf{r}_{2}-\mathbf{r}_{i}| &= \frac{R(t_0) (\lambda_i - \mu_i)}{2}.
\end{align}
For the initial condition of bound electron $i$ to be physically meaningful, its kinetic energy must be non-negative. Thus, we require $P_i \geq 0$, where
\begin{equation}
P_i = 2\left[ -I_{p,2}-W_i\left(\lambda_3, \mu_3, \phi_3,\lambda_4, \mu_4, \phi_4\right) \right].
\end{equation}
Therefore, the final form of the microcanonical distribution is 
\begin{align}\label{eq:microcanonical}
\begin{split}
&\tilde{\rho}\left(\lambda_3, \mu_3, \phi_3,\lambda_4, \mu_4, \phi_4\right) \\
&=\left\{\begin{array}{ll}
\prod_{i=3}^4 \left(\lambda_i^2- \mu_i^2\right) \sqrt{P_i} & \text { for all }  P_i \geq 0 \\
0 & \text { for any } P_i<0.
\end{array}\right.
\end{split}
\end{align}
Next, to generate initial conditions for O$_2^{2+}$ we need to identify the range of values of $\lambda_i,\mu_i,\phi_i$ so that $P_i \geq 0$ for each bound electron $i$. We find that $\mu_i\in [-1,1]$ and $\phi_i\in [0,2 \pi]$ for each electron. That is,  $P_i \geq 0$ is satisfied for the whole range of values of $\phi_i$ and $t_i.$ In addition, for $P_i \geq 0$ to be satisfied we find that $\lambda_i$ cannot be larger than $\lambda_{\text{max}},$ i.e.,  $\lambda_i\in [1,\lambda_{\text{max}}].$ The value $\lambda_{\text{max}}$ is the same for both bound electrons. For this range of values, then, we find the maximum value $\tilde{\rho}_{\text{max}}$ of the microcanonical distribution $\tilde{\rho}\left(\lambda_3, \mu_3, \phi_3,\lambda_4, \mu_4, \phi_4\right)$ given in \eq{eq:microcanonical}. Next, we generate the uniform random numbers $\lambda_i\in [1,\lambda_{\text{max}}]$, $\mu_i\in [-1,1]$, $\phi_i\in [0,2 \pi]$ for each electron, and $\chi \in [0,\tilde{\rho}_{\text{max}}]$. If  $\tilde{\rho}\left(\lambda_3, \mu_3, \phi_3,\lambda_4, t_4, \phi_4\right)>\chi$ then the generated values of $\lambda_i, t_i$ and $\phi_i$ are accepted as initial conditions, otherwise, they are rejected and the sampling process starts again.

Once we find  $\lambda_i, \mu_i$ and $\phi_i,$ we obtain the position vector $\mathbf{r}_{i} = (r_{x,i},r_{y,i},r_{z,i})$ and the momentum vector $\mathbf{p}_{i} = (p_{x,i},p_{y,i},p_{z,i})$ of each electron $i$ as follows
\begin{align}
r_{x,i} &= \frac{R(t_0) \cos \left( \phi_i \right)}{2} \sqrt{ \left( \lambda_i^2 - 1\right)\left( 1 - \mu_i^2\right)  }\\
r_{y,i} &= \frac{R(t_0) \sin \left( \phi_i \right)}{2} \sqrt{ \left( \lambda_i^2 - 1\right)\left( 1 - \mu_i^2\right)   }\\
r_{z,i} &= \frac{R(t_0) \lambda_i  \mu_i}{2} \\
p_{x,i} &= \sqrt{P_i} \cos \left( \phi_{\mathbf{p},i} \right) \sqrt{1 - \nu_{\mathbf{p},i}^2}\\
p_{y,i} &=\sqrt{P_i} \sin \left( \phi_{\mathbf{p},i} \right) \sqrt{1 - \nu_{\mathbf{p},i}^2}\\
p_{z,i} &=\sqrt{P_i} \nu_{\mathbf{p},i},
\end{align}
where $\phi_{\mathbf{p},i} \in [0,2 \pi]$ and $\nu_{\mathbf{p},i} \in [-1,1]$ define the momentum $\mathbf{p}_i$ in spherical coordinates. 

The potential energy employed in the microcanonical distribution does not include the dipole interaction term, $- Q_j\mathbf{r}_{j} \cdot \mathbf{E}\left(\mathbf{r}_{j}, t\right),$ which appears in the energy expression used to determine the effective charges, see \eq{eq:energy_of_electron_j}. 
 As a consequence, once the initial conditions of the two bound electrons are sampled, the corresponding effective charges $\zeta_{j,n}(t_0),$ must be recalculated using \eq{eqn::zeta_and_energy}, which does include the dipole interaction. The energies of the electrons are coupled due to the term $C_{j,n}(\mathcal{E}_{j},|\mathbf{r}_{1}-\mathbf{r}_{j}|,|\mathbf{r}_{2}-\mathbf{r}_{j}|)$. Therefore, we obtain the effective charges via an iterative procedure. If the energies, and thus the effective charges, do not converge at $t_0$, the initial conditions are rejected and re-sampled.  We note that for the laser intensities considered in this work, the percentage of rejected trajectories is negligible.

\subsection{Tunneling during propagation}\label{Sec::Tunnelling}
During time propagation, we allow for each bound electron to tunnel at the classical turning points along the axis of the electric field using the Wentzel-Kramers-Brillouin (WKB) approximation \cite{WKB}. The transmission probability is given by \cite{WKB},
\begin{equation}\label{EQ:Notrootz}
T \approx \exp \left(-2 \int_{r_a}^{r_b}\left[2\left(V_{\mathrm{tun}}\left(r, t_{\mathrm{tun}}\right)-\epsilon_i\right)\right]^{1 / 2} d r\right),
\end{equation}
with $V_{\mathrm{tun}}\left(r, t_{\mathrm{tun}}\right)$ the potential of the bound electron i in the presence of the nuclei and the laser field, including also the effective potential terms, and $r$ the distance of the electron along the field \cite{PhysRevA.109.033106}.  The energy of electron $i$ at the time of tunneling, $t_{\text{tun}}$, is $\epsilon_i$. The classical turning point are denoted as $r_a$ and $r_b$. 

Here, we have added an extra condition for tunnelling to take place during time propagation compared to Ref. \cite{PhysRevA.109.033106}. Specifically, when an electron tunnels, its position changes which in turn results in a change in  the energies and the effective charges of the other bound electrons. This results in a change in the energy of the tunneling electron and so on.  Hence, we allow for tunneling during propagation to occur only if the new position of the tuneling electron  leads to  the new energies of all bound electrons being convergent. Accounting for tunneling during time propagation is necessary  to accurately describe phenomena related to enhanced ionization \cite{Enhanced1,Enhanced2,Enhanced3,Enhanced4,Enhanced5} during the fragmentation of strongly driven molecules.

Finally, we note that we treat tunnelling differently at the start and during time propagation. The reason is that at the start to initiate the $t_0$ time, we need a tunnelling rate, since the Monte Carlo method, part of the ECBB model, integrates over different $t_0$ times  in the laser pulse. However, once the propagation starts, we no longer need a rate but a tunneling probability instead.  We choose to employ the one given by the WKB approximation \cite{WKB}.

\section{Results}
Using the ECBB model for molecules, we focus on triple ionization (TI), frustrated triple ionization (FTI), double ionization (DI), and frustrated double ionization (FDI) in the three-active-electron O$_2^+$. The molecule is driven by an 800 nm, 40 fs laser pulse at two low intensities of  0.5 and 1 $\mathrm{PW/cm^2}$ as well as two high intensities of 5 and 7 $\mathrm{PW/cm^2}$. We  consider two orientations of the molecule with respect to the laser pulse, namely,   the two nuclei being parallel or perpendicular to the direction of the electric field, i.e., along the $z$ or $x$ axis, respectively.  

In triple ionization, three electrons escape and two O$_2^{2+}$  ions are formed. In frustrated triple ionization, two electrons escape and one electron finally remains bound at a Rydberg state of either of the two ion frargments. In double ionization, two electrons escape and the O$_2^{2+}$ and O$_2^{+}$  ions are formed. In frustrated double ionization, only one electron escapes and the process leads to the formation of two O$_2^{+}$ ions, or to the formation of one O$_2^{2+}$ ion and a neutral O. We find the former channel of FDI to be significantly more probable compared to the latter, and,  hence, we do not address in this work the latter.

Finally,  as in our previous studies, see for example Ref. \cite{PhysRevA.109.033106}, we identify the principal quantum number $n$ for each Rydberg electron by first calculating the classical principal quantum number
\begin{equation}
n_c = \frac{Q_{c}}{\sqrt{2|\epsilon_i(t_f)|}}, 
\end{equation}
with $\epsilon_i(t_f)$ being the energy of a bound electron at the end of the time propagation, and $Q_{c}$ the charge of the nucleus where the electron remains bound. Then, we assign a quantum number $n$ so that the following criterion is satisfied \cite{Comtois_2005},
\begin{equation}
\left[  \left( n-1\right)\left( n-\frac{1}{2}\right)n \right]^{1/3} \leq n_c \leq \left[  n\left( n+\frac{1}{2}\right)\left( n+1\right) \right]^{1/3}.
\end{equation}

\subsection{Probabilities}
\begin{figure}[b]
\centering
\includegraphics[width=\linewidth]{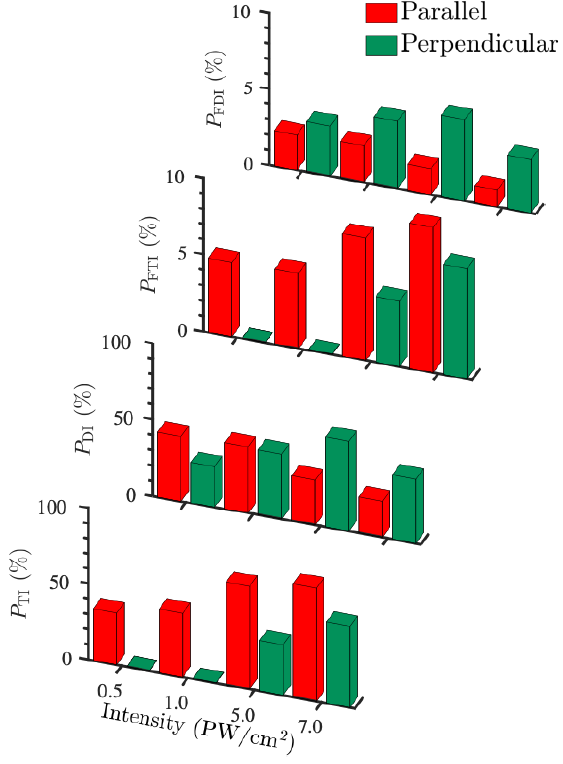}
\caption{Probabilities of the TI, FTI with $n>2$, DI and FDI with $n>2$, for  O$_2^+$, for the four intensities and two molecular orientations considered in this work.}\label{Fig:Probabilities}
\end{figure}

In \fig{Fig:Probabilities}, we plot the probabilities of the main ionization processes in O$_2^+$ as a function of  intensity and molecular orientation. For parallel orientation, we find that TI becomes increasingly dominant with increasing intensity, rising from 34\% at 0.5 $\mathrm{PW/cm^2}$ to roughly 74\% at 7 $\mathrm{PW/cm^2}$. In contrast, for the perpendicular orientation, DI is the prominent ionization channel for the three lower intensities, reaching roughly 60\% at 5 $\mathrm{PW/cm^2}$. At the highest intensity, TI takes over DI for the perpendicular orientation as well. Also, we find that DI significantly contributes for  both orientations, whereas TI is strongly suppressed at lower intensities for the perpendicular orientation. 
 Frustrated ionization pathways occur with significantly lower probabilities compared to their non-frustrated counterparts. However, they exhibit a similar dependence on both intensity and orientation. In particular, FTI is strongly suppressed for the perpendicular alignment at 0.5 and 1 $\mathrm{PW/cm^2}$. In contrast, FDI is enhanced for the perpendicular alignment and exhibits a weak but consistent increase with increasing intensity until 5 $\mathrm{PW/cm^2}$ before dropping slightly at 7 $\mathrm{PW/cm^2}$.

\subsection{Kinetic energy release distributions}\label{Sec:Results_KER}

\begin{figure}[t]
\centering
\includegraphics[width=\linewidth]{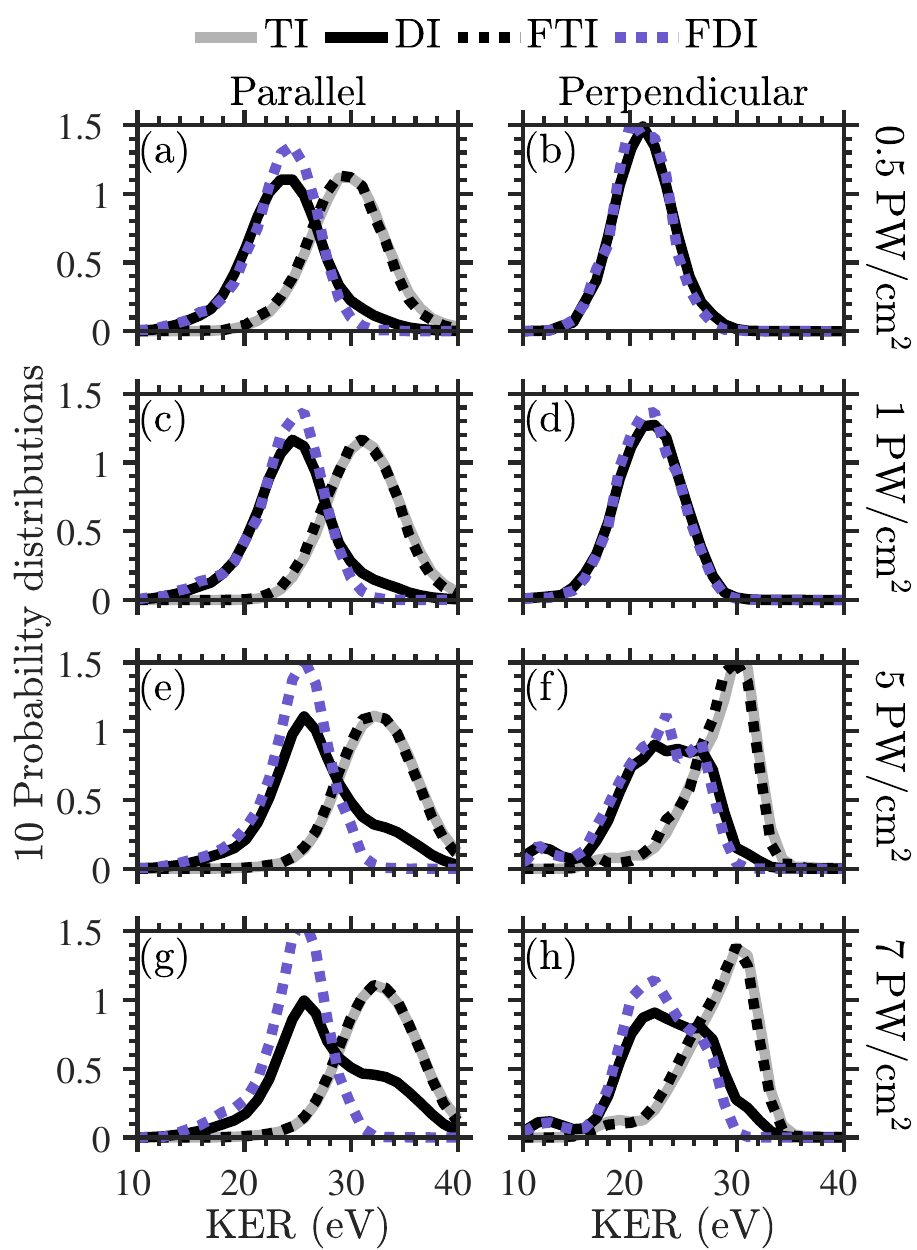}
\caption{Distribution of the sum of the final kinetic energies of the ions produced in triple ionization (gray solid lines), double ionization (black solid lines), frustrated triple ionization (black dotted lines) and frustrated double ionization (gray dotted lines) for $\mathrm{O_2^+}$ oriented parallel (a),(c),(e),(g) and perpendicular (b),(d),(f),(h) to the electric field. All distributions are normalized to one.}\label{Fig:KER}
\end{figure}

In \fig{Fig:KER}, we plot the kinetic energy release (KER) distributions of the final ion fragments for triple ionization, double ionization, frustrated triple ionization, and frustrated double ionization at the four laser intensities and for both molecular orientations of O$_2^+$. For the perpendicular orientation at the two lowest intensities, the probabilities of TI and FTI are negligible, therefore, the corresponding KER distributions are not presented in \fig{Fig:KER}. In all cases, we find the KER distributions for TI and FTI to be nearly identical. This behaviour is consistent with the Rydberg electron in FTI remaining bound in a highly excited state and, hence, not screening significantly the nuclei. The DI and FDI distributions are similar, mostly for small intensities, while at higher ones FDI has on average smaller values. The main reason is that FTI is also registered as DI, giving rise 
to the bump in the KER at higher energies for higher intensities. In FDI, in addition to the Rydberg electron, one electron remains deeply bound in an $n=1$ state, resulting in an increased screening of the nuclei and, hence, in smaller final kinetic energies of the ion fragments. Also, as expected, we find (not shown) that the sum of the kinetic energies of the nuclei plus the Coulomb repulsion of the nuclei at the time the last electron ionizes correspond to the peak of the final KER distributions. Finally, at the two higher intensities of 5 $\mathrm{PW/cm^2}$ [Figs. \ref{Fig:KER}(e) and \ref{Fig:KER}(f)] and 7 $\mathrm{PW/cm^2}$ [Figs. \ref{Fig:KER}(g) and \ref{Fig:KER}(h)] all ionization processes shift toward higher KER. This is consistent with the increased intensity of the laser field leading to Coulomb explosion of the nuclei setting in at earlier times compared to smaller intensities.

\begin{figure}[t]
\centering
\includegraphics[width=\linewidth]{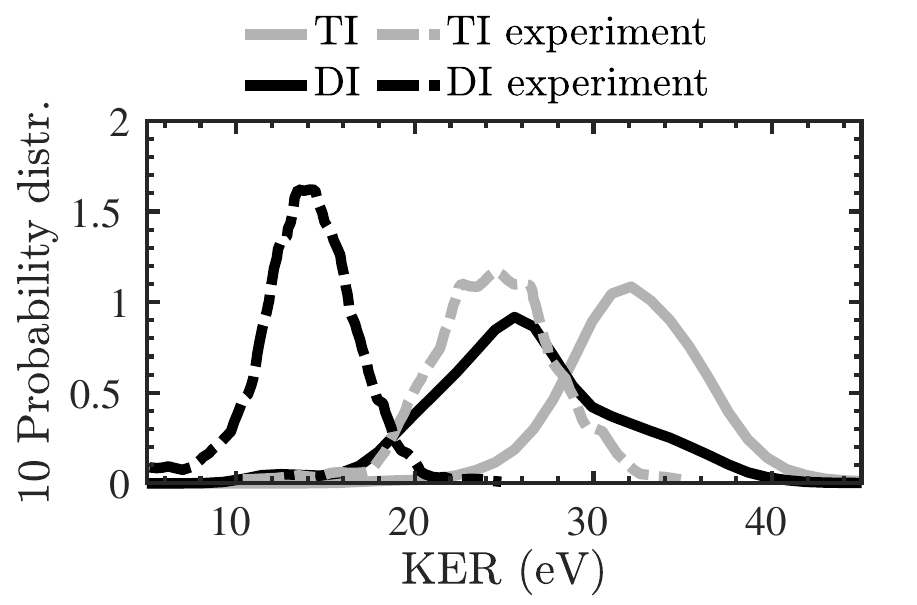}
\caption{Distribution of the sum of the final kinetic energies of the ions produced in triple ionization (gray solid lines) and double ionization (black solid lines) at 7 $\mathrm{PW/cm^2}$ averaged over both orientations. We also include experimental distributions from Ref. \cite{PhysRevA.79.063414} for triple ionization (gray dotted lines), double ionization (black dotted lines). All distributions are normalized to one.}\label{Fig:KER_exp}
\end{figure}

In \fig{Fig:KER_exp}, we  compare at 7 $\mathrm{PW/cm^2}$  the KER obtained with the ECBB model  with experiment \cite{PhysRevA.79.063414}.  We plot the KER for TI and DI when averaged over both orientations. We find the results from the ECBB model to be consistently larger compared to the experiment for both double and triple ionization. Indeed, the theoretical KER distribution for TI (solid gray line) peaks at roughly 32 eV, while the experimental one peaks at approximately 24 eV. For DI, the theoretical KER distribution (solid black line) is broad and peaks at 25 eV while the experimental distribution is narrow and peaks at 14 eV. 

To understand this difference, we focus on the impact of all the forces on the kinetic energies of the two nuclei, that is,  we split up the final momentum of each nucleus into the momentum changes from every different force
\begin{align}\label{Eq:core_mom}
\begin{split}
&\mathbf{p}_i(t_f) = \mathbf{p}_i(t_0) + \Delta \mathbf{p}^{ C, \mathrm{rep} }_i(t_0 \to t_f) + \Delta \mathbf{p}^{ C, \mathrm{att} }_i(t_0 \to t_f)\\
& + \Delta \mathbf{p}^{ \mathbf{E} }_i(t_0 \to t_f) + \Delta \mathbf{p}^{ \mathbf{B} }_i(t_0 \to t_f) + \Delta \mathbf{p}^{ V_{ \mathrm{eff} } }_i(t_0 \to t_f), 
\end{split}
\end{align}
with $i=1,2$. The nuclei are initiated at rest, hence, $\mathbf{p}_i(t_0)=0$. Also, we find the momentum change due to the electric and the magnetic field, $\Delta \mathbf{p}^{ \mathbf{E} }_i$ and $\Delta \mathbf{p}^{ \mathbf{B} }_i$ respectively, to be negligible. Hence, we  focus on the remaining terms, i.e., the momentum changes from the Coulomb repulsion between nuclei, Coulomb attraction between electrons and nuclei and effective potentials that are expressed through the nuclei in the approximation adopted in the ECBB model, denoted as, $ \Delta \mathbf{p}^{ C, \mathrm{rep} }_i, \Delta \mathbf{p}^{ C, \mathrm{att} }_i, \Delta \mathbf{p}^{ V_{ \mathrm{eff} } }_i$ respectively. As expected, we find  the $z$ component of the momentum of the nuclei to be the one contributing the most to the KER, and only  focus  on this component.
 
For all ionization processes, we find that  the nuclei move in opposite directions along the $z$ axis, with nucleus 1 moving to the left ($-z$ axis) and nucleus 2 to the right $+z$ axis. We find that the Coulomb repulsion between the two nuclei during molecular fragmentation contributes the most to the momenta of the nuclei,  i.e. $ \Delta p^{ C, \mathrm{rep} }_{z,1}<0$ and $ \Delta p^{ C, \mathrm{rep} }_{z,2}>0$ have the largest values  in \eq{Eq:core_mom}. 

The Coulomb attraction each nucleus experiences from the bound electrons has  the opposite effect, i.e., it attracts each nucleus towards the origin. Indeed,  prior to Coulomb explosion at time $t_{ \mathrm{frag}}$, the bound electrons move in between the two nuclei. Hence, the Coulomb attraction on average leads each nucleus towards the origin, resulting to a positive $ \Delta p^{ C, \mathrm{att} }_{z,1}(t_0 \to t_{ \mathrm{frag}} )$ and a negative $ \Delta p^{ C, \mathrm{att} }_{z,2}(t_0 \to t_{ \mathrm{frag} } )$. After  fragmentation, if an electron remains bound in a nucleus $n$, $ \Delta p^{ C, \mathrm{att} }_{z,n}(t_{ \mathrm{frag}} \to t_f)$ will average to zero. However, if it remains bound in the other nucleus, then it will attract nucleus $n$ towards the origin leading to $ \Delta p^{ C, \mathrm{att} }_{z,1}(t_{ \mathrm{frag}} \to t_f)>0$ and $ \Delta p^{ C, \mathrm{att} }_{z,2}(t_{ \mathrm{frag}} \to t_f)<0$. Also, for TI and DI, we find that the biggest contribution to  $ \Delta p^{ C, \mathrm{att} }_{z,n}(t_0 \to t_f)$ stems from the electron that ionizes last, since it interacts for a longer time with the nuclei, while the smallest contribution stems from the electron that ionizes first. 

However, in our model there in an extra force acting on the nuclei due to our replacing of the full Coulomb potential with effectives ones to describe the interaction between bound electrons. From \eq{eq:eff_potential}
\begin{align*}
&V_{\text{eff}}(\zeta_{i,n}(t),|\mathbf{r}_{n}-\mathbf{r}_{j}|)  = \frac{1 - (1+\zeta_{i,n}|\mathbf{r}_{n}-\mathbf{r}_{j}|)e^{-2\zeta_{i,n}|\mathbf{r}_{n}-\mathbf{r}_{j}|}}{|\mathbf{r}_{n}-\mathbf{r}_{j}|},
\end{align*}
one can see that the potential electron $j$ experiences due to electron $i$ depends also on the position vector of the nucleus, $\mathbf{r}_{n}.$ Hence, when a pair of bound electrons $(i,j)$ interact via the effective potential, the momentum of the nuclei changes artificially. To better understand this momentum change, we calculate the force acting on the nucleus due to the effective potential $V_{\text{eff}}(\zeta_{i,n}(t),|\mathbf{r}_{n}-\mathbf{r}_{j}|) $
\begin{align}\label{Eq:Force_Veff}
\begin{split}
&\mathbf{F}(\zeta_{i,n},| \mathbf{r}_{n} - \mathbf{r}_{j} |) =-\dfrac{\partial V_{\text{eff}}(\zeta_{i,n},| \mathbf{r}_{n} - \mathbf{r}_{j} |)}{\partial \mathbf{r}_n}\\
&=-\dfrac{-1+\left[ 1+2\zeta_{i,n} | \mathbf{r}_{n} - \mathbf{r}_{j} |(1+\zeta_{i,n} | \mathbf{r}_{n} - \mathbf{r}_{j} |)\right] }{| \mathbf{r}_{n} - \mathbf{r}_{j} |^3} \\
& \times e^{-2\zeta_{i,n} | \mathbf{r}_{n} - \mathbf{r}_{j} |} (\mathbf{r}_{n} - \mathbf{r}_{j}).
\end{split}
\end{align}
 From \eq{Eq:Force_Veff}, one can readily see that if $z_n-z_j<0$ then the force and hence, $\Delta p^{ V_{ \mathrm{eff} } }_{z,n}$ is positive. Similarly, if $z_n-z_j>0$, the momentum change is negative. As previously discussed, before fragmentation, the bound electrons are on average moving between the two nuclei, leading to $z_1-z_j<0$ and $z_2-z_j>0$. Hence, the momentum change due to the effective potential will be negative for nucleus 1 and positive for nucleus 2, assisting fragmentation. We expect the change in momentum of the nuclei due to the effective potential to be larger for DI compared to TI. Indeed, in TI no electrons remain bound, and  we find the momentum change of the nuclei due to the effective potential to be small. The largest contribution stems from the effective potential of the two bound electrons that ionize last.  In contrast, in DI,  we have one electron that finally remains bound. In DI, we find that the  effective potential that contributes the most to the momentum change of a nucleus $n$   originates from the effective force that an electron $i$ that is bound on the other nucleus exerts on an electron $j$ bound to the nucleus $n$, i.e., from the term $V_{\text{eff}}(\zeta_{i,n}(t),|\mathbf{r}_{n}-\mathbf{r}_{j}|)$. Again, this is positive for nucleus 1, since $z_1-z_j<0$ and negative for nucleus 2 since  $z_2-z_j > 0.$ Hence, overall, the effective forces lead to larger to larger final momenta of the ion fragments and hence  higher KER.  This is more so for processes where at least one electron remains bound. 
 
\begin{figure}[t]
\centering
\includegraphics[width=\linewidth]{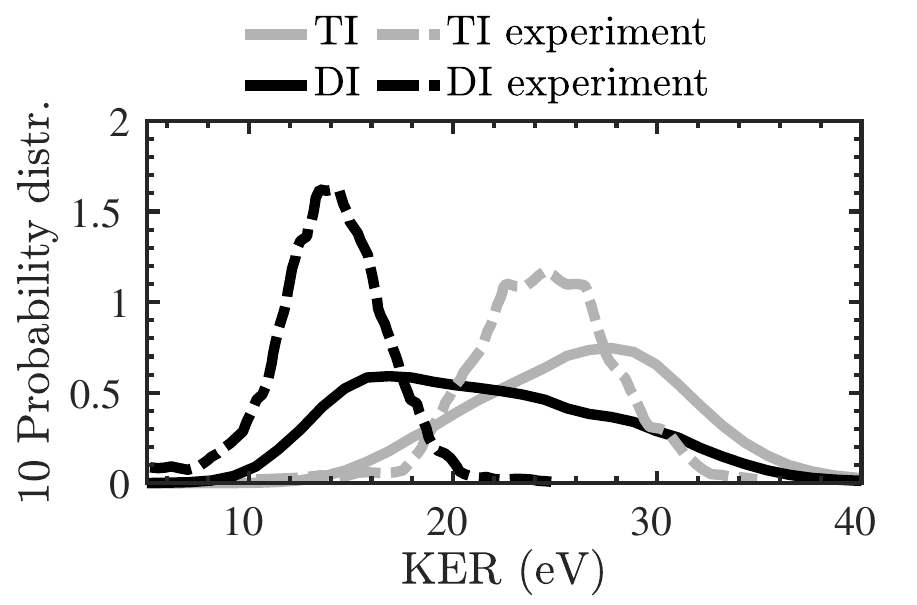}
\caption{Distribution of the sum of the final kinetic energies of the ions produced in triple ionization (gray solid lines) and double ionization (black solid lines) at 7 $\mathrm{PW/cm^2}$ averaged over both orientations, calculated with $\Delta \mathbf{p}^{ V_{ \mathrm{eff} } }_i(t_0 \to t_f)$ excluded in \eq{Eq:core_mom}. We also include experimental distributions from Ref. \cite{PhysRevA.79.063414} for triple ionization (gray dotted lines), double ionization (black dotted lines). All distributions are normalized to one.}\label{Fig:KER_aprox}
\end{figure}

  The above are  consistent with the DI spectra obtained with the ECBB model having a larger difference from experiment compared to TI, see \fig{Fig:KER_exp}. Indeed,  in \fig{Fig:KER_aprox}, 
 we plot  the KER when the momentum change of the nuclei due to the effective potentials is excluded. We now find a much better agreement between the results obtained with the ECBB model and experiment, especially for TI.
 
 The above discussion also reveals when the effective potentials currently employed by the ECBB model lead to results closer to experiment for the KER. We expect that the KER obtained with the ECBB model will be more accurate  the smaller the force on the nuclei is due to the effective potential between two bound electrons.  Indeed, for multi-centre molecules, the probability for a nucleus not to have an electron bound increases, which in turn decreases the force via the effective potential on this nucleus. Hence, for the same number of electrons, we expect that the KER results will be closer to experiment the more atoms the molecule  has. Also, we expect that the KER closest to experiment will be for the ionization process where all electrons escape, since for this process the effective potentials have the smallest effect.

\subsection{Correlation between ionization times and internuclear distance at the time of ionization}\label{Sec:Results_R_vs_tion}

\begin{figure}[H]
\centering
\includegraphics[width=\linewidth]{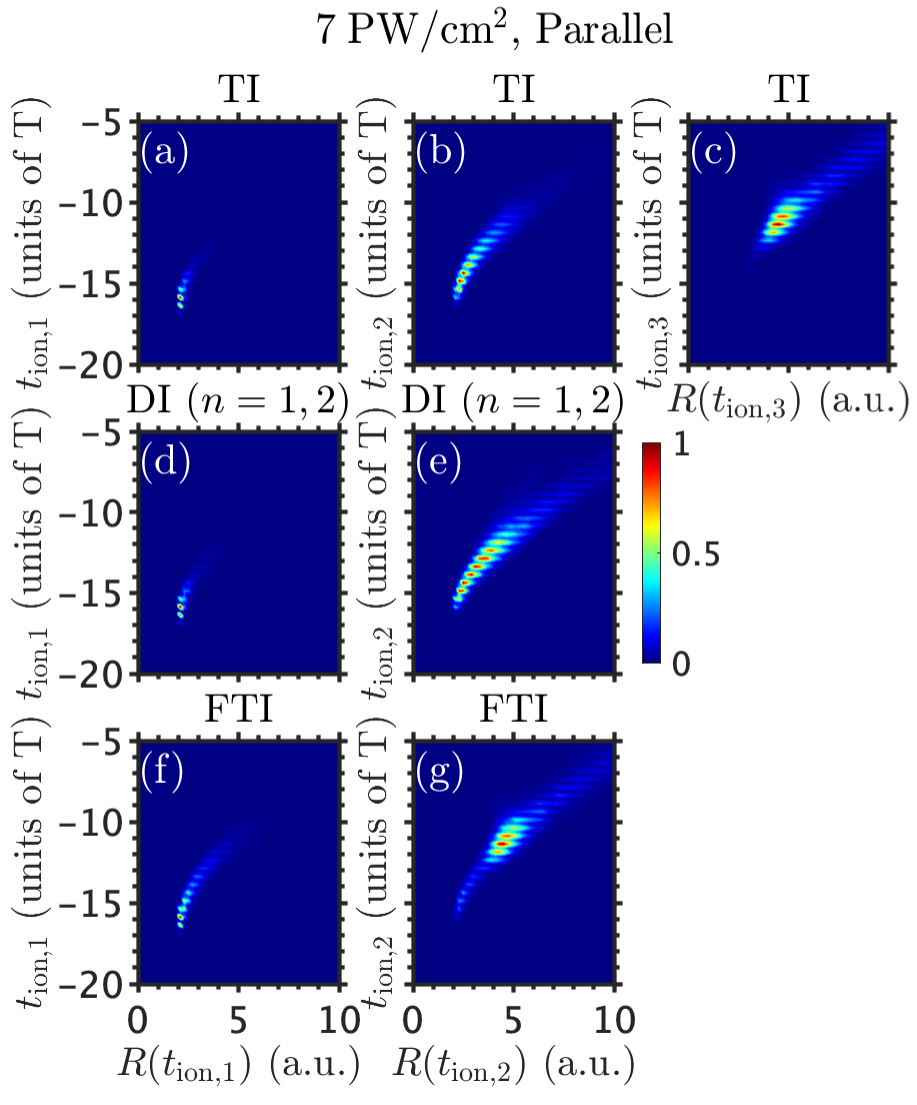}
\caption{Distributions of the correlated ionization times and internuclear distances at the ionization time for every ionizing electron in TI (a),(b),(c), DI with $n=1,2$ (d),(e) and FTI with $n>2$ (f), (g).  All distributions are normalized to one.}\label{Fig:R_vs_tion}
\end{figure}

In \fig{Fig:R_vs_tion},  for TI, DI and FTI at 7 $\mathrm{PW/cm^2}$ and parallel orientation, we plot   the correlation between the ionization times $t_{ \text{ion},i }$ and the internuclear distance at the time of ionization $R \left( t_{ \text{ion},i } \right) $. The index $i$ ranges from 1 to 3, with 1 corresponding to the electron that ionizes first and 3 to the last.  For DI, we plot only the events where the remaining bound electron occupies an $n=1, 2$ state, since these are the ones not included in FTI.

For TI  [Figs. \ref{Fig:R_vs_tion}(a)-\ref{Fig:R_vs_tion}(c)], we find that the first electron ionizes early at small internuclear distances. The second electron ionizes at  larger distances, albeit smaller that the critical distance for enhanced ionized to occur. This indicates a correlated electron escape of the two fastest-to-ionize electrons. Indeed, we find that in roughly 65\% of the TI events, there is a recollision  between the first- and the second-to-ionize electrons, with the latter escaping shortly after. The third electron ionizes at later times and larger internuclear distances, consistent with ionization driven predominantly by enhanced ionization. 
We find a similar behaviour for DI [Figs. \ref{Fig:R_vs_tion}(d)-\ref{Fig:R_vs_tion}(e)]. In DI compared to TI,  the second electron ionizes at larger internuclear distances, still however smaller than for distances consistent with enhanced ionization.  

In contrast to DI, for FTI [Figs. \ref{Fig:R_vs_tion}(f)-\ref{Fig:R_vs_tion}(g)] we find a different behaviour for the second-to-ionize electron. Indeed, comparing \fig{Fig:R_vs_tion}(g) with \fig{Fig:R_vs_tion}(c) we find that the second ionization of FTI closely resembles the third ionization of TI. This suggests that FTI proceeds as follows. The first electron tunnel-ionizes early on and returns later to the molecular ion for a recollision, as in DI and TI. Following recollision, the returning electron gets re-captured in a excited state while one of the bound electrons gains energy and ionizes.  This is consistent with the ionization time versus internuclear distance in the first ionization of FTI resembling mostly the second ionization of TI. This can be better understood from the discussion in Sec. \ref{Sec:Paths}, where we show that pathway B of FDI prevails at the highest intensity.
Finally, the last-to-ionize electron, does so via enhanced ionization, as in TI.

\subsection{Pathways of frustrated ionization}\label{Sec:Paths}
\begin{figure}[H]
\centering
\includegraphics[width=\linewidth]{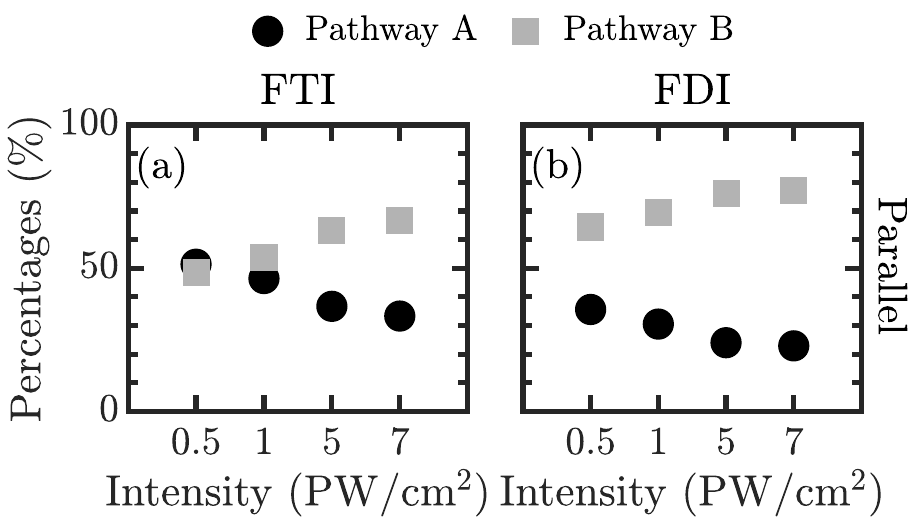}
\caption{Percentages of pathways A and B for FTI and FDI for all intensities for the parallel and perpendicular orientation.}\label{Fig:Pathways}
\end{figure}
We find that FTI and FDI proceed via two pathways, first identified in the FDI of the strongly driven two-electron molecule $\mathrm{H_2}$ \cite{toolkit2014}, and then identified in FTI and FDI of the strongly driven three-electron molecule $\mathrm{HeH_{2}^+}$ \cite{PhysRevA.109.033106}. In both pathways, we find that the initially tunnelling electron ionizes early on (first step), while one of the remaining bound electrons does so later in time (second step). If the second (first) ionization step is frustrated, we label the  pathways as A and B, respectively.

In \fig{Fig:Pathways}, we plot the percentage contributions of pathways A and B to the formation of frustrated triple ionization and frustrated double ionization as a function of laser intensity for parallel orientation. We find that at lower intensities both pathways contribute roughly the same for the formation of FTI, with pathway B becoming more important with increasing intensity. Indeed, at 7 $\mathrm{PW/cm^2}$ the percentage contribution of pathway B is approximately double the one of pathway A. This trend is more apparent for FDI, with pathway B being the prevalent one even at lower intensities and its contribution increasing  as the intensity increases. 

\begin{figure}[H]
\centering
\includegraphics[width=\linewidth]{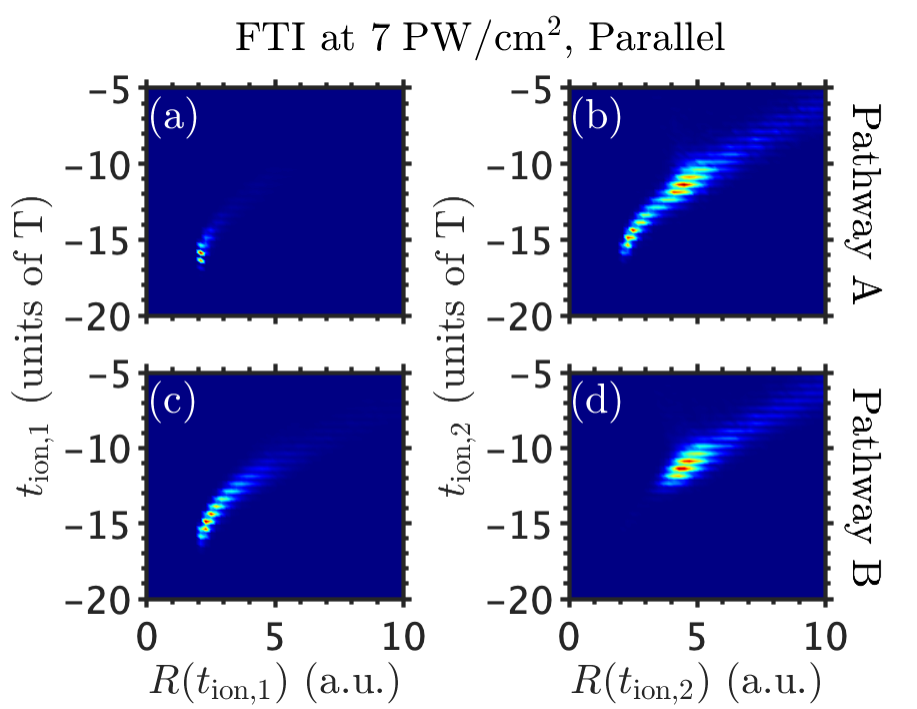}
\caption{Distributions of the correlated ionization times and internuclear distances at the ionization time for every ionizing electron in pathway A [(a), (b)] and pathway B [(c), (d) ] of FTI with $n>2$ .  All distributions are normalized to one.}\label{Fig:R_vs_tion_pathways}
\end{figure}

Pathway B being the dominant one in FTI at high intensity is consistent with our findings in the previous section. Namely, in \fig{Fig:R_vs_tion}(f), the times of ionization for the first ionizing electron in FTI versus the internuclear distance resembles the second-to-ionize electron in \fig{Fig:R_vs_tion}(b) of TI.  Indeed, since pathway B prevails in FTI, this means that the tunnel-ionizing electron which returns to recollide gets finally captured in a Rydberg state, while it transfers enough energy to one of the bound electrons to finally escape. It is mostly this latter electron plotted in \fig{Fig:R_vs_tion}(f) of FTI, resembling the second-to-ionize electron in  \fig{Fig:R_vs_tion}(b) of TI, the difference being that the returning electron ionizes in TI while it does not in FTI. The above become more clear in \fig{Fig:R_vs_tion_pathways} which show the ionization times of each electron versus the internuclear distance  separately for pathway A and B of FTI. We clearly see that \fig{Fig:R_vs_tion_pathways}(a) for the first-to-ionize electron in pathway A of FTI  resembles \fig{Fig:R_vs_tion}(a) for the first-to-ionize electron in TI . However, \fig{Fig:R_vs_tion_pathways}(c) for the first-to-ionize electron in pathway B of FTI resembles \fig{Fig:R_vs_tion}(b) for the second-to-ionize electron in TI.

\begin{figure}[b]
\centering
\includegraphics[width=\linewidth]{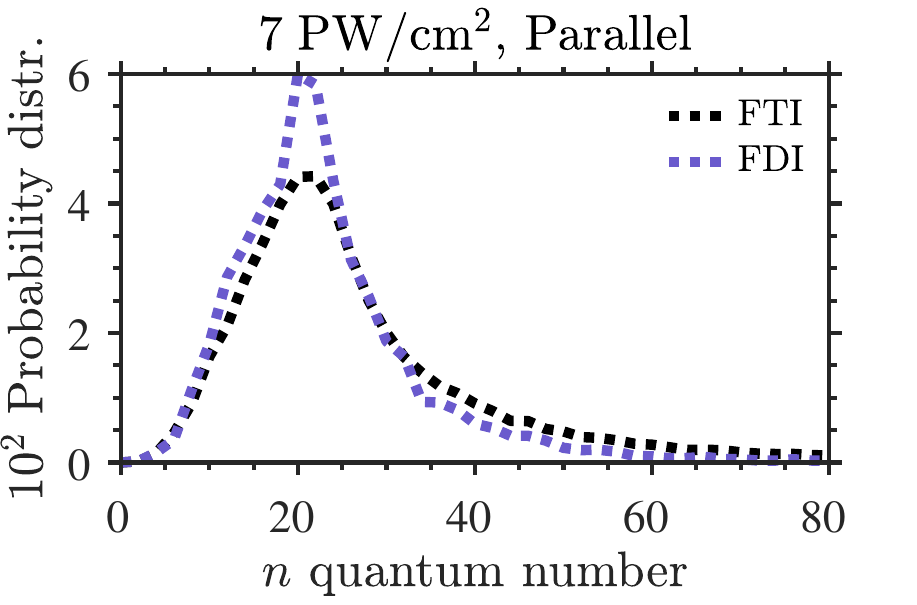}
\caption{Distribution of the principal quantum number $n$ of the Rydberg electrons for FTI (black dotted line) and FDI (purple dotted line). Both distributions are normalized to one.}\label{Fig:Quantum_number} 
\end{figure}

Finally, in \fig{Fig:Quantum_number} we investigate the distribution of the principal $n$ quantum number for FTI  and FDI at an intensity of 7 $\mathrm{PW/cm^2}$ and for parallel orientation. Similar results are found for the other intensities and orientation.  We find that both distributions are similar, both peaking around $n\approx 20$ while exhibiting long tails extending to large quantum numbers, with the distribution for FTI extending to slightly higher $n$ numbers. This is expected. In contrast to FDI, in FTI we only have one electron remaining bound, leading to higher nuclear charge. Assuming the Rydberg electron has the same energy in FDI and FTI, higher nuclear charge in FTI  means higher $n$ numbers.  We find (not shown) that the distribution of $n$ numbers is roughly the same for  pathways A and B.

\section{Conclusions}
We have extended a recently developed three-dimensional semiclassical model to study correlated multielectron ionization during the fragmentation of $\mathrm{O_2^+}$ driven by intense, infrared laser pulses at various intensities and for two orientations of the diatomic molecule with respect to the electric field. We focused on the main ionization processes, namelly, triple, double, frustrated triple and frustrated double ionization. We calculated the sum of the final kinetic energies of all ion fragments and compared with available experimental results at 7 $\mathrm{PW/cm^2}.$ We found that our model overestimates the KERs, especially for double ionization. This disagreement between theory and experiment stems mainly from the artificial increase of the final momentum of the cores, due to the effective potential used to describe the interaction between bound electrons. These results serve
 as an example of the current limitations of the ECBB model as well as a starting point for future improvements. They also suggest that, for a given number of electrons,  the KERs with the ECBB model will be more accurate the more atomic centres a molecule has. Furthermore, we investigated the dynamics of multi-electron escape for ionization and frustrated processes. For triple and double ionization, the first two electrons to ionize are correlated undergoing a soft recollision, while in TI the last electron ionizes mainly due to enhanced ionization. For frustrated triple ionization at high intensities the electron that remains bound in a Rydberg state is mainly the electron returning to recollide. The electron that ionizes first is mainly the bound electron that gains energy from the returning electron, while the last electron ionizes via enhanced ionization as in triple ionization.  Our findings demonstrate that the ECBB model is a powerful tool in understanding multi-electron escape in  molecules driven by intense, infrared laser pulses.

\section{Acknowledgements}
The authors acknowledge the use of the UCL Myriad High Performance Computing Facility (Myriad@UCL), the use of the UCL Kathleen High Performance Computing Facility (Kathleen@UCL), and associated support services in the completion of this work.

\appendix
\section{Ionization rate}\label{Sec::ionrate}
We find $t_0$ using as the sampling distribution the tunnel ionization rate derived  in Ref. \cite{PhysRevLett.106.173001}
\begin{equation}\label{eq:App12}
		\Gamma=2\pi \kappa^2C^2_\kappa\Bigg(\frac{2\kappa^3}{E(t_0)}\Bigg)^{ \frac{2Q}{\kappa} -1}\exp\Bigg(-\frac{2\kappa^3}{3E(t_0)}\Bigg)R(\theta_L),
	\end{equation}
where $E(t_0)$ is the instantaneous field strength, $\theta_L$ is the angle between the laser field and the $z$ axis in the lab frame, $\kappa=\sqrt{2I_{p1}}$ and $Q$ is the asymptotic charge which is equal to two for O$_2^+.$  We find the coefficient $C_\kappa$ by fitting the Dyson orbital \cite{Patchkovskii2007} to the following asymptotic form of the wave function \cite{PhysRevLett.106.173001}
	\begin{equation}\label{eq:App13}
		 \Psi(\mathbf{r})\approx C_\kappa\kappa^{3/2}(r_M\kappa)^{Q/\kappa-1}e^{-\kappa r_M}F(\theta_M,\phi_M),
	\end{equation}
where $r_M,\theta_M$ and $\phi_M$ are the spherical coordinates in the molecular frame. The Dyson orbital \cite{Patchkovskii2007} is the overlap integral of the wave function of the ground state of O$_2^+$
\begin{equation}
\left(1 \sigma_g^2, 1 \sigma_u^2, 2 \sigma_g^2, 2 \sigma_u^2, 3 \sigma_g^2, 1 \pi_{u x}^2, 1 \pi_{u y}^2, 1 \pi_{g x}^1 \right)
\end{equation}
 with the ground state of O$_2^{2+}$ computed at the equilibrium distance of O$_2^+$
\begin{equation}
\left(1 \sigma_g^2, 1 \sigma_u^2, 2 \sigma_g^2, 2 \sigma_u^2, 3 \sigma_g^2, 1 \pi_{u x}^2, 1 \pi_{u y}^2 \right).
\end{equation}
Both wave functions were found using MOLPRO \cite{MOLPRO_brief} with the Hartree-Fock method. 

The function $F(\theta_M,\phi_M)$ depends on the outer-valence molecular orbital the electron occupies before tunnel-ionizing at the start of the propagation.  An expression for $F$ is provided in \cite{PhysRevLett.106.173001,Radzig1985}
\begin{align}
\begin{split}
F(\theta_M,\phi_M) =& \cosh \left( \frac{\kappa R(t_0)}{2} \cos \theta_M  \right) \left(1 + \alpha  \cos ^2\theta_M \right) \\
&\times\cos \theta_M  \sin \theta_M \cos \phi_M.
\end{split}
\end{align}
We fit the Dyson orbital in the intervals $r_{\mathrm{min}} \le r_M \le r_{\mathrm{max}}$, $0\le\theta_M\le\pi$ and $0\le\phi_M\le2\pi$ to find the parameters $C_\kappa$ and $\alpha$. The interval for $r_M$ was chosen so that for $r_M>r_{\mathrm{min}}$ the Coulomb potential corresponding to O$_2^{2+}$ has effectively the form of a one-center Coulomb potential, i.e. $-Q/r$; the upper limit was chosen so that  for $r_M>r_{\mathrm{max}}$ the Dyson orbital is practically zero.

As discussed in \cite{PhysRevLett.106.173001} (shown also here for completeness), the function $R(\theta_L)$ is given by
\begin{align}
\begin{split}\label{EQ:6}
R\left(\theta_L\right)= & {\left[F_0\left(\theta_L\right)-\frac{4E\left(t_0\right)}{3 \kappa^3} F_2\left(\theta_L\right)+\frac{2E\left(t_0\right)}{3 \kappa^3} F_3\left(\theta_L\right)\right]^2 }\\
&+\frac{2E\left(t_0\right)}{9 \kappa^3} F_1^2\left(\theta_L\right)
 \end{split}
\end{align}
where the functions $F_j(\theta_L)$ are obtained using the following formulas \cite{PhysRevLett.106.173001}
\begin{align}
\begin{split}\label{eq:App16}
F_0(\theta_L)&=F(u,v), \\ 
F_1(\theta_L)&=F_v\cos\theta_L-F_u\sin\theta_L, \\ 
F_2(\theta_L)&=F_u\cos\theta_L+F_v\sin\theta_L, \\
F_3(\theta_L)&=F_{vv}\cos^2\theta_L+F_{uu}\sin^2\theta_L-F_{uv}\sin2\theta_L,
\end{split}
\end{align}
where we set $u = \cos \theta_M $ and $v = \sin \theta_M \cos \phi_M.$ $F_u$, $F_{v}$, $F_{uu}$, $F_{vv}$, and $F_{uv}$ are the first and second order partial derivatives of $F(u,v)$ with
respect to $u$ and $v$, calculated at $u=\cos\theta_L$ and $v=\sin\theta_L$.

\bibliography{bibliography}{}

\end{document}